

\magnification\magstep1
\hoffset=4pt
\parindent=1.5em
\TagsOnRight
\baselineskip=12pt
\NoBlackBoxes
\font\ninepoint=cmr9

 \NoBlackBoxes

meros reales
meros enteros

\def\k{\kappa}                 
\def\kp{\k_1}                  
\def\kl{\k_2}                  





do
do
do

do

            %
            %
           %
          %
          %

\redefine\C{\text{\ \!C}}      
\redefine\S{\text{\ \!S}}      %
\define\>#1{{\bold#1}}                 
n para vectores

\define\co{\Delta}                     

\define\conm#1#2{\left[#1,#2 \right]} 
\def\1{\'{\i}}                         

\define\q#1{{\left[#1\right]}_q}      
\define\qq#1{{\left\{ #1\right\} }_q}  

\font\cabeza=cmbx12

\centerline{\cabeza CAYLEY-KLEIN LIE ALGEBRAS AND THEIR}
\smallskip
\centerline{\cabeza QUANTUM UNIVERSAL ENVELOPING
ALGEBRAS\footnote{Communication
 presented in the ``International Symposium on Non Associative Algebras and
Applications", Oviedo (Spain), July 1993.}}
\bigskip
\smallskip

\centerline {A. Ballesteros, F.J. Herranz, {M.A. del Olmo} and M.
Santander} \bigskip \centerline{\it Departamento de F\1sica Te\'orica,
Universidad de Valladolid.}

\centerline{\it E-47011, Valladolid. Spain.}
\smallskip

\bigskip

\ninepoint
\noindent
ABSTRACT. The N-dimensional Cayley-Klein scheme allows
the simultaneous description of $3^N$ geometries (symmetric orthogonal
homogeneous spaces) by
 means of a set of Lie algebras depending on $N$ real parameters.  We
present here a  quantum deformation of the Lie algebras generating the
groups of motion of the two and  three dimensional Cayley-Klein geometries.
This deformation (Hopf algebra structure) is presented in a compact form by
using a formalism developed for the case of  (quasi)free Lie algebras.
Their quasitriangularity (i.e., the most usual way to study the
associativity of their dual objects, the quantum groups) is also discussed.

\rm
\bigskip\bigskip

\noindent\cabeza 1. Introduction\rm
\bigskip

We study here a certain type of Lie algebra deformations (so called
``quantum" ones), that have recently appeared in the context of the Quantum
Inverse Scattering Method. They are properly defined as deformations of the
corresponding universal enveloping algebra $U \frak g$ and their dual
objects (in certain restricted sense) generate the ``quantum groups"
--deformations of the algebra of functions on the group, in the spirit of
non-commutative geometry [1,2]--. Their underlying algebraic structure
(mainly Hopf algebra properties [3]) is rather rich and was soon described
for the classical simple Lie algebras [4,5,6].

\medskip

However, many physically interesting groups are not simple groups: for
instance,  the groups of inertial transformations of space--time such as
Galilei or Poincar\'e ones. Some quantum deformations have been built for
their associated Lie algebras  [7, 8] which also arise as symmetries of
certain physical problems [9]. We present  here an attempt towards their
characterization based on a Cayley-Klein (CK) geometrical
 scheme that includes all these groups as well as their transformed by
In\"on\"u--Wigner contractions [10]. We also discuss the problems arising in
the definition of the quantum groups as dual objects of these quantum
algebras, mainly in connection with the way in which the associativity of
the deformed algebra of functions on the group is guaranteed (the
$R$--matrix problem).

\bigskip

\noindent\cabeza 2. The Cayley-Klein Lie algebras\rm \medskip

{}From a physical point of view, some interesting homogeneous symmetric
spaces can be simultaneously described in the framework of CK geometries.
To do this, we consider an $N$-dimensional ($N$-d) symmetric orthogonal
geometry as a group $G$
 of dimension $\tfrac 1 2 N(N+1)$ and a set of $N$ commuting involutions
$S^{(i)}$  in $\frak g$, the Lie algebra of $G$. If we denote by $\frak
h^{(i)}$ the Lie
 subalgebras of elements invariant under $S^{(i)}$, their corresponding
groups $H^{(i)}$
 have to be taken as the isotropy groups of a point, a line, \dots a
$(N-1)$-flat, so the homogeneous spaces $\Cal X^{(i)}\equiv G/H^{(i)}$ turn
out to be the spaces of points, lines, \dots of the geometry. The
involutions must satisfy certain requirements (specially on the dimensions
of the subgroups $H^{(i)}$) and the group $G$ is also required to act
effectively on all the
 $\Cal X^{(i)}$. Without entering into details, the main result is that the
CK Lie  algebra depends on $N$ real parameters $(\k_1,  \dots,\k_N)$ that
can take any real value (however, these parameters can be scaled to $-1,\,
0$ or $+1$), and we have [11].

\medskip

\it Let $G$ be the CK Lie group corresponding to an $N$-d CK geometry.

\item{a)} The Lie algebra of $G$ has dimension $\tfrac 1 2 N(N+1)$ with
generators $J_{ij}$ $(i<j;\, i,j=0,1,\dots,N)$ and is characterized
 by $N$ real parameters $(\k_1, \dots,\k_N)$.

\item{b)} The isotopy subgroups of $i$-flats $H^{(i)}$ $(i=0,1,\dots,N-1)$,
correspond to the subalgebras $\frak h^{(i)}=\langle J_{ab},J_{cd} \rangle$
$(a<b,\, a,b=0,1,\dots,i;\ c<d;\,\, c,d=i+1,\dots,N )$.

\item{c)} The Lie brackets of the basis elements $J_{ij}$ can be written in
terms of the parameters $\k_{ij}$ $(i<j;\, i,j=0,1,\dots,N)$ defined by

$\k_{ij}=\k_{i+1}\k_{i+2}\dots\k_{j},$ and are
\vskip-6pt
$$ [J_{ij},J_{lm}]
=\delta_{im}J_{lj}-\delta_{jl}J_{im}+\delta_{jm}\k_{lm}J_{il}+
\delta_{il}\k_{ij}J_{jm},\quad (i\leq l,\ j\leq m). \tag 2.1
$$

\item{d)} The CK algebras $\frak g_{(\k_1,\dots ,\k_N)}$ can be realized in
terms of $(N+1) \times (N+1)$ real matrices: \vskip-6pt $$ \Cal
D(J_{ij})=-\k_{ij}e_{ij}+e_{ji},\tag 2.2 $$ where $e_{ij}$ are the standard
$(N+1) \times (N+1)$ matrices with elements
$(e_{ij})_{kl}=\delta_{ik}\delta_{jl}$ and commutation relations
\vskip-10pt
$$
[e_{ij},e_{lm}]=\delta_{jl}e_{im}-\delta_{im}e_{lj}. \tag 2.3
$$
\rm

\medskip

A representation of the oneparameter subgroups associated to the generators
 $J_{ij}$ is obtained by exponentiation of the matrices $\Cal D(J_{ij})$.
The matrix
 entries of these subgroup elements are easily written by using the
``generalized"
 trigonometric  functions $\text{sine}_{\k}(x)\equiv\S_{\k}(x)$ and

$\text{cosine}_{\k}(x)\equiv\C_{\k}(x)$,  defined in terms of power series
as follows:
$$
\S_{\kappa}(x)=\sum_{l=0}^{\infty} (-\kappa)^l
{x^{2l+1}\over  (2l+1)!}\ ,\quad \C_{\kappa}(x)= \sum_{l=0}^{\infty}
(-\kappa)^l {x^{2l}\over  (2l)!}\ . \tag2.4
$$
These functions constitute
an intrinsic tool throughout any explicit computation within the CK scheme
(either classical or quantum), and can be considered as deformations of the
``Galilean" or ``parabolic" functions $1$ and $x$, to which they tend in
the limit $\k\to 0$:
$$
\C_{\k}(x) = \cases \cos {\sqrt{\k} x}
\qquad\quad \text{if }\k>0 \\  1 \qquad \qquad\qquad \text{if } \k=0 \\
\cosh {\sqrt{-\k} x} \quad\ \  \text{if }\k<0 \\  \endcases ,\quad\
\S{_\k}(x) = \cases \frac{1}{\sqrt{\k}} \sin {\sqrt{\k} x} \qquad\quad\

\text{if }\k>0 \\ x \qquad\qquad\qquad\quad\ \ \text{if } \k=0 \\
\frac{1}{\sqrt{-\k}} \sinh {\sqrt{-\k} x} \quad \ \text{if }\k<0 \\
\endcases  .\tag 2.5
$$
\vfill\eject
\bigskip

\noindent\cabeza 3. Quantum Universal Enveloping algebras\rm
\medskip

Following Drinfel'd [4], a ``quantized universal enveloping algebra"
 (QUE algebra) of a Lie algebra $\frak g$ is a Hopf algebra $\Cal A$ over
the formal
 power series $\Bbb C [[z]]$ on a deformation indeterminate $z$ such that
$\Cal A$  is topologically  free $\Bbb C [[z]]$--module and $\Cal A /z\Cal
A$ is isomorphic  (as Hopf algebra) to  $U\frak g$.

\medskip

Recall that a $\Bbb C$--algebra is a Hopf algebra if there exist two
 homomorphisms  called coproduct $(\Delta : \Cal A \longrightarrow \Cal A
\otimes \Cal A )$ and counit  $(\epsilon : \Cal A \longrightarrow \Bbb C)$,
as well as an antihomomorphism (the antipode $\gamma : \Cal A
\longrightarrow \Cal A$) such that, $\forall a\ \! \in  \Cal A$:
$$
\align
(id\otimes\co)\co (a)&=(\co\otimes id)\co (a),   \tag 3.1\\
(id\otimes\epsilon)\co (a)&=(\epsilon\otimes id)\co (a)= a,  \tag 3.2\\
m((id\otimes \gamma)\co (a))&=m((\gamma \otimes id)\co (a))= \epsilon (a)
1,  \tag 3.3
\endalign
$$
where $m$ is the usual multiplication $m(a\otimes
b)=ab$. Counit and antipode are derived from the coproduct in a unique way.

\medskip

Roughly speaking, we can think of the elements in $\Cal A$ as formal power

series in  $z$ with coefficients in $U\frak g$. Provided we are not going
to take  into account topological properties, the topological freeness is
translated into  algebraic terms by imposing that the QUE algebra must be
free, or at least torsion--free
 as a $\Bbb C [[z]]$--module [12]. Both the Hopf homomorphisms and the
torsion condition restrict the number of formal power series suitable for
quantization [13].

\medskip

The ``classical limit" property $\Cal A /z\Cal A \simeq U\frak g$ is also
 a very  strong constrain on the quantization. Our aim is to obtain quantum
CK algebras, so the $z\rightarrow 0$ limit must lead to the Lie brackets
described in (2.1). The standard quantization of the classical Cartan
series of simple Lie algebras uses the $A_1$ quantum structure as building
block [4,5]. For the quantum CK algebras already obtained [14,15], the
analogous generalization seems not to be immediate. However, CK geometries
of a given dimension do contain lower dimension subgeometries. Any compact
way of writing of the Hopf homomorphisms embodying somehow this embedding
would help to obtain the general structure. In this sense, the following
proposition [16] ensures the consistency of the deformation for a
``quasi"--free algebra (the only condition is the commutativity of the
primitive generators) and will be used to write the quantum CK algebras in
the next section.

\medskip

\proclaim{Proposition 3.1} Let $\{1, H_1,\dots,H_n,X_1,\dots,X_m\}$ be the
generators of  a ``free" (up to the conditions $\conm{H_i}{H_j}=0\
\,\forall i,j$) associative algebra $E$ over $\Bbb C$ . Let
$\alpha_i,\beta_j\ \, i,j=1,\dots,n$ be a set of $(m\times m)$ matrices
with  entries in $\Bbb C [[z]]$ such that
$\conm{\alpha_i}{\beta_j}=\conm{\alpha_i}{\alpha_j}=\conm{\beta_i}{\beta_j}=0
\ \,\forall i,j$. Let $\vec X$ be a ``vector" with components $X_l,\ \,
 l=1,\dots,m$. The coproduct
$$
\aligned \co 1 &=1 \otimes 1,\\ \co H_i &=1
\otimes H_i + H_i\otimes 1,\\ \co\vec X &= \exp (\sum_{i=1}^n{\alpha_i
H_i}) \dot \otimes \vec X +
 \sigma\left( \exp (\sum_{i=1}^n{\beta_i H_i}) \dot\otimes \vec X \right),
\endaligned \tag 3.4
$$
turns the completion $B$ of formal power series on
$z$ with coefficients in  $E$ into a Hopf algebra. \endproclaim

We have denoted by $\sigma$ the permutation map $\sigma (a\otimes
b)=b\otimes a$ and, if $P\equiv (p_{kl})$ is a $(m\times m)$ matrix with
entries in $B$,  the $k$-th component of $(P \dot\otimes \vec X)_k$ is
defined as
\vskip -6pt
$$
(P\dot\otimes \vec X)_k=\sum_{l=1}^m
p_{kl}\otimes X_l. \tag 3.5
$$

\bigskip

\noindent\cabeza 4. QUE Cayley-Klein algebras\rm
\medskip

We restrict in this section to the study of the algebras generating the two
and three dimensional CK systems. The former is a family of 3-d Lie algebra
depending on two parameters which contains as particular cases
 $so(3),so(2,1)$, the 2-d euclidean $e(2)$ and the $(1+1)$ Newton-Hooke,
Galilei and Poincar\'e algebras. A simultaneous quantization of all of
these systems  is given in the following theorem.

\medskip

\proclaim{Theorem 4.1} Let $\frak g_{(\kp,\kl)}$ be the Lie algebra
 generating the 2-d CK systems and whose infinitesimal generators are
$\{J_{12},P_1,P_2\}$. The coproduct
$$
\aligned
\co P_2 &=1 \otimes P_2 +
P_2\otimes 1,\\ \co\pmatrix P_1 \\ J_{12} \endpmatrix &= \exp\left\{
\pmatrix  -\tfrac z 2 P_2 & 0 \\ 0 & -\tfrac z 2 P_2 \endpmatrix \right\}

\dot\otimes \pmatrix P_1 \\  J_{12} \endpmatrix + \sigma\left( \exp\left\{
\pmatrix \tfrac z 2 P_2 & 0 \\ 0 & \tfrac z 2 P_2 \endpmatrix \right\}
\dot\otimes \pmatrix P_1 \\  J_{12} \endpmatrix  \right), \endaligned \tag
4.1
$$
and the commutation relations
 $$
\conm{{J}_{12}}{{ P}_1}=\S_{-z^2}
\left({ P}_2\right), \quad  \conm{{J}_{12}}{{ P}_2}= -\k_2 { P}_1, \quad

\conm{{P}_{1}}{{ P}_2}= \k_1 { J}_{12}. \tag 4.2
$$
define the QUE algebra $U_z\frak g_{(\kp,\kl)}$.
\endproclaim

\medskip

\proclaim{Corollary 4.1} Counit and antipode are deduced from (4.1) and,
for $X\in \{P_1, P_2,J_{12}\}$, read
$$
\epsilon(X) =0,\qquad\qquad
\gamma(X)=-e^{{z\over 2}  P_2}\ X\ e^{-{z\over 2}  P_2}. \tag 4.3
$$
\endproclaim

\proclaim{Corollary 4.2} The center of $U_z\frak g_{(\kp,\kl)}$ is
generated by
$$
 C_z=4 \C_{\k_1\k_2}({\tfrac z 2})\left[ \S_{-z^2}({\tfrac 1 2} P_2)
\right]^2 + \tfrac 2 {z} \S_{\k_1\k_2}({\tfrac z 2}) \left\{ \k_2 P_1^2  +
\k_1 J_{12}^2\right\} . \tag 4.4
$$
\endproclaim

\proclaim{Proposition 4.1} The fundamental representation $D_q$ of $U_z\frak
g_{(\kp,\kl)}$ in terms of the ``classical" one $D$ is defined as follows
$$
D_q(P_2)=D(P_2),\quad D_q(P_1)=\sqrt{{{\S_{\k_1\k_2}(z)}\over{z}}}
D(P_1),\quad D_q(J_{12})=\sqrt{{{\S_{\k_1\k_2}(z)}\over{z}}} D(J_{12}).
\tag 4.5
$$
\endproclaim

\medskip

Note that the ``generalized trigonometric functions" appear as natural
 deformation functions in this context. Moreover, they are consistent
formal power  series in the sense that, for instance, $\co
(\S_\k(P_2))=\S_\k(P_2)\otimes\C_\k(P_2) + \C_\k(P_2)\otimes\S_\k(P_2)$.
The classical limit $z\rightarrow 0$ is  always well defined and
straightforwardly leads to the algebra defined in (2.1). It is also worth
remarking that different algebras are obtained by specialization of the
$\k_i$ parameters. This deformation preserves always non--Lie character in
(4.2), since the deformed bracket cannot vanish whatever the $\k_i$ are.

\medskip

For the case of 3-d geometries, the CK Lie algebra is now 6-d and depends
on three measure coefficients $(\k_1,\k_2,\k_3)$. The number of different
geometries included is now $3^3$. The algebras  $so(4),so(3,1),so(2,2)$,
the 3-d Euclidean algebra $e(3)$, the $(2+1)$--d versions of the
Newton-Hooke, Galilei and Poincar\'e algebras can be found among the set
$\frak g_{(\k_1,\k_2,\k_3)}$.

\medskip

\proclaim{Theorem 4.2} Let $\frak g_{(\k_1,\k_2,\k_3)}$ be the Lie algebra
 generating the 3-d CK systems. We denote the infinitesimal generators as
$J_{ij}$, $(i<j;\, i,j=0,1,2,3)$. The coproduct
$$
 \aligned
\co(J_{03})&=1\otimes J_{03} + J_{03}\otimes 1,\qquad\qquad
\co(J_{12})=1\otimes J_{12} + J_{12}\otimes 1,\\  \co\pmatrix J_{01} \\
J_{02} \\ J_{13} \\ J_{23}\endpmatrix &=  \exp\left\{ \alpha_1 J_{03} +
\alpha_2 J_{12} \right\}  \dot\otimes \pmatrix J_{01} \\  J_{02} \\ J_{13}
\\ J_{23}  \endpmatrix + \sigma\left( \exp\left\{ \beta_1 J_{03} + \beta_2
J_{12} \right\} \dot\otimes \pmatrix J_{01} \\  J_{02} \\ J_{13} \\ J_{23}
\endpmatrix  \right), \endaligned \tag 4.6
$$
$$
\alpha_1=-\beta_1=\pmatrix
-\tfrac z 2 & 0 & 0 & 0  \\
 0 & -\tfrac z 2 & 0 & 0 \\ 0 & 0 & -\tfrac z 2 & 0  \\ 0 & 0 & 0 & -\tfrac
z 2 \endpmatrix, \qquad  \alpha_2=-\beta_2=\pmatrix 0 & 0 & 0 & -\tfrac z 2
\k_1  \\
 0 & 0 & \tfrac z 2 \k_1  & 0 \\ 0 & \tfrac z 2 \k_3 & 0 & 0  \\ -\tfrac z
2 \k_3 & 0 & 0 & 0 \endpmatrix,
$$
and the following non--vanishing
commutation relations
$$
\aligned [J_{12},J_{01}]&=J_{02},\qquad
[J_{12},J_{02}]=-\k_2J_{01},\qquad
[J_{01},J_{02}]=\k_1\S_{-z^2\k_1\k_3}(J_{12})\C_{-z^2}(J_{03}),\\
[J_{13},J_{01}]&=\S_{-z^2}(J_{03})\C_{-z^2\k_1\k_3}(J_{12}),\qquad
[J_{13},J_{03}]=-\k_2\k_3J_{01},\quad [J_{01},J_{03}]=\k_1J_{13},\\
[J_{23},J_{02}]&=\S_{-z^2}(J_{03})\C_{-z^2\k_1\k_3}(J_{12}),\qquad

[J_{23},J_{03}]=-\k_3J_{02},\qquad  [J_{02},J_{03}]=\k_1\k_2J_{23} ,\\

[J_{23},J_{12}]&=J_{13},\qquad
[J_{23},J_{13}]=-\k_3\S_{-z^2\k_1\k_3}(J_{12})\C_{-z^2}(J_{03}), \qquad
[J_{12},J_{13}]=\k_2J_{23}. \endaligned \tag4.7
$$
define the QUE algebra
$U_z\frak g_{(\k_1,\k_2,\k_3)}$.
\endproclaim

\medskip

\proclaim{Corollary 4.3} Counit and antipode are
$$ \epsilon(J_{ij})
=0,\qquad\qquad \gamma(J_{ij})=-e^{{z}  J_{03}}\ J_{ij}\ e^{-{z}  J_{03}}.
\tag 4.8
$$
\endproclaim

\proclaim{Corollary 4.4} The following deformed second order elments belong
to the center of $U_z\frak g_{(\kp,\kl,\k_3)}$
$$
\aligned \Cal C_1^q &= 4
\C_{\k_{03}}(z) \left[\S_{-z^2}^2(\tfrac 1 2
J_{03})\C_{-z^2\k_1\k_3}^2(\tfrac 1 2 J_{12}) + \k_1\k_3 \S_{-z^2
\k_1\k_3}^2(\tfrac 1 2 J_{12})\C_{-z^2}^2(\tfrac 1 2 J_{03})\right] \\
&\qquad\quad + \tfrac 1 z \S_{\k_{03}}(z)\left[\k_2\k_3 J_{01}^2 + \k_3
J_{02}^2 + \k_1 J_{13}^2 + \k_1\k_2 J_{23}^2 \right],\\ \Cal C_2^q &=
\C_{\k_{03}}(z) \S_{-z^2}(J_{03}) \S_{-z^2\k_1\k_3}(J_{12}) + \tfrac 1 z
\S_{\k_{03}}(z) [ \k_2 J_{01} J_{23} - J_{02} J_{13}]. \endaligned \tag 4.9
$$
\endproclaim

\proclaim{Proposition 4.2} The fundamental representation $D_q$ of
$U_z\frak g_{(\kp,\kl,\k_3)}$ is given by
$$
\aligned
 D_q(J_{ij})&=\sqrt{\tfrac 1 {z}\S_{\k_{03}}(z)}\, D(J_{ij}),\quad
{\text{if}} \ ij=01,02,13,23,\\
 D_q(J_{ij})&=   D(J_{ij}),\qquad\qquad\qquad {\text{if }} \ \! ij=03,12.
\endaligned \tag4.10
$$
\endproclaim

\bigskip

\noindent\cabeza 5. Quasitriangular Hopf algebras\rm \medskip

A quasitriangular Hopf algebra [4] is a pair $(\Cal A,\Cal R)$ where
$\Cal A$ is a Hopf algebra and $\Cal R \in \Cal A\otimes \Cal A$ is
invertible and obeys
$$
\aligned \sigma \circ \co h = \Cal R (\co h) \Cal
R^{-1}, &\qquad \forall\ \, h \in  \Cal A \\ (\co \otimes id)\Cal R =\Cal
R_{13}\Cal R_{23},\qquad & (id \otimes \co) \Cal R =\Cal R_{13}\Cal R_{12},
\endaligned \tag 5.1
$$
where, if $\Cal R=\sum_{i} a_i\otimes b_i$, we
denote $\Cal R_{12}\equiv\sum_{i} a_i\otimes b_i\otimes 1$,  $\Cal
R_{13}\equiv\sum_{i} a_i\otimes 1\otimes b_i$ and $\Cal
R_{23}\equiv\sum_{i} 1\otimes a_i\otimes b_i$. If $\Cal A$ is a
 quasitriangular Hopf algebra, then  $\Cal R$ is called an ``universal"
$\Cal R$--matrix  and satisfies the  {\it Quantum Yang--Baxter Equation}:
$$
\Cal R_{12}\Cal R_{13}\Cal R_{23}=\Cal R_{23}\Cal R_{13}\Cal R_{12}. \tag
5.2
$$

Given a matrix representation $\rho:\Cal A\rightarrow Mat(n,\Bbb C)$, the
matrix  elements $t_{ij}$ of its dual $\Cal A^\ast$ (the ``quantum group")
satisfy the
 commutation relations
 $$
 R T_1 T_2= T_2 T_1 R, \tag 5.3
$$
where $R=(\rho\otimes\rho)(\Cal R)$ and $T=(t_{ij})$, $T_1=T\otimes 1_n$ and
$T_2=1_n\otimes T$. These matrix solutions $R$ are relevant for physical
applications. If $R$ satisfies (5.2), this equation ensures that the

non--commutative algebra generated by the $t_{ij}$ is associative (third
order  commutation relations are derived from second order ones). However,
for non--semisimple  quantum algebras (in our scheme, if some $\k_i=0$),
neither the universal $\Cal R$ matrices nor even some of their particular
realizations are easy to find [7,17]. In fact, they could not exist.

\bigskip \noindent {\bf Acknowledgments.} This work has been partially
supported by  DGICYT (Spain) under project PB92--0255.

\bigskip\bigskip

\eightpoint

\noindent\bf References \rm
\eightpoint

\smallskip

\ref     
\no[{1}]
\by      Yu I. Manin \jour    {\sl Quantum groups and non--commutative
geometry}. CRM, Montreal
\vol
\yr       1989
\pages
\endref

\ref     
\no[{2}]
\by       S.L. Woronowicz
\jour     Comm. Math. Phys.
\vol      111
\yr      1987
\pages   613

\moreref
\paper II
\jour     Invent. Math,
\yr 1988
\vol 93
\pages 35
\endref

\ref
\no[3]
\by         E. Abe
\book       Hopf Algebras
\publ       Cambridge Tracts in Mathematics 74,
\publaddr   Cambridge University Press, Cambridge
\yr         1980
\endref

\ref      
\no[{4}]
\by V.\, G.\, Drinfeld
\paper Quantum Groups
\jour Proceedings of the International Congress of Mathematics,  MRSI
Berkeley, (1986) 798
\endref

\ref      
\no[{5}]
\by M. Jimbo
\jour Lett. Math. Phys.
\yr 1985
\vol 10
\pages 63

\moreref \paper \    \yr 1986 \vol 11 \pages 247 \endref

\ref     
\no[{6}]
\by      N. Yu. Reshetikhin, L.A. Takhtadzhyan and L.D. Faddeev
\jour    Leningrad Math. J.
\vol     1
\yr      1990
\pages   193
\endref

\ref
\no[{7}]
\by      E. Celeghini, R. Giachetti, E. Sorace and M. Tarlini
\paper   Contractions of quantum groups
\jour    Lecture Notes in Mathematics n. 1510, 221, (Springer-Verlag, 1992)

\endref

\ref     
\no[{8}]
\by       J. Lukierski, H. Ruegg and A. Nowicky
\jour     Phys. Lett. B.
\vol      293
\yr       1992
\pages    344
\endref

\ref     
\no[{9}]
\by       F. Bonechi, E. Celeghini, R. Giachetti, E. Sorace and M. Tarlini

\jour     Phys. Rev. Lett.
\vol      68
\yr      1992
\pages   3178
\endref

\ref       
\no[{10}]
\by        E. In\"on\"u, E. P. Wigner
\jour      Proc. Natl. Acad. Sci. U. S.
\vol       39
\yr        1953
\pages     510
\endref

\ref        
\no[{11}]
\by         M. Santander, F.J. Herranz, M.A. del Olmo
\book       Proceedings of the  XIX ICGTMP
\publ       Anales de F\1sica, Monograf\1as. Vol. 1.I, p. 455.
CIEMAT/RSEF
\publaddr   Madrid (1993)
\yr
\endref

\ref
\no[{12}]
\by S.\, Majid
\jour Int. J. Mod. Phys. A
\vol 5 \yr 1990 \pages 1
\endref

\ref \no[{13}]
\by P. Truini and V.S. Varadarajan
\jour Rev. Math. Phys.
\vol 5 \yr 1993 \pages 363
\endref

\ref        
\no[{14}]
\by         A. Ballesteros, F.J. Herranz, M.A. del Olmo and M. Santander
\paper
\jour       J. Phys. A: Math. Gen.
\vol        26
\pages        5801 \
yr         1993
\endref

\ref        
\no[{15}]
\by         A. Ballesteros, F.J. Herranz, M.A. del Olmo and M. Santander
\paper     {\sl Quantum (2+1) kinematical algebras: a global approach}.
\jour       J. Phys. A: Math. Gen.
\vol        27
\yr         1994
\endref

\ref \no[{16}]
\by V. Lyakhovsky and A. Mudrov
\jour J. Phys. A: Math. Gen.
\vol 25
\yr 1992 \pages 1139
\endref

\ref \no[{17}]
\by A. Ballesteros, E. Celeghini, R. Giachetti, E. Sorace and M. Tarlini
\jour J. Phys. A: Math. Gen.
\vol 26 \yr 1993 \pages 7495
\endref

\end